\documentclass[runningheads]{llncs}
\usepackage[T1]{fontenc}
\usepackage{graphicx}
\usepackage{url}
\usepackage{placeins}
\usepackage[raggedrightboxes]{ragged2e}

\begin{document}
\title{Detecting streaks in smart telescopes images with Deep Learning}
\titlerunning{Detecting streaks in smart telescopes images with Deep Learning}

%
\author{Olivier Parisot\inst{1}\orcidID{0000-0002-3293-3628}, Mahmoud Jaziri\inst{1}\orcidID{0009-0008-6350-1423} }
\authorrunning{O. Parisot, M. Jaziri}

\institute{
	Luxembourg Institute of Science and Technology (LIST) \\ 
	5 Avenue des Hauts-Fourneaux, 4362 Esch-sur-Alzette, Luxembourg \\ 
	\email{olivier.parisot@list.lu}
}

\maketitle              
\begin{abstract}
The growing negative impact of the visibility of satellites in the night sky is influencing the practice of astronomy and astrophotograph, both at the amateur and professional levels. The presence of these satellites has the effect of introducing streaks into the images captured during astronomical observation, requiring the application of additional post processing to mitigate the undesirable impact, whether for data loss or cosmetic reasons. In this paper, we show how we test and adapt various Deep Learning approaches to detect streaks in raw astronomical data captured between March 2022 and February 2023 with smart telescopes.

\keywords{Astronomy \and Streaks detection \and Deep Learning.}
\end{abstract}

\section{Introduction}
\label{sec:intro}

\begin{figure}[h]
	\centering
	\includegraphics[width=.9\linewidth]{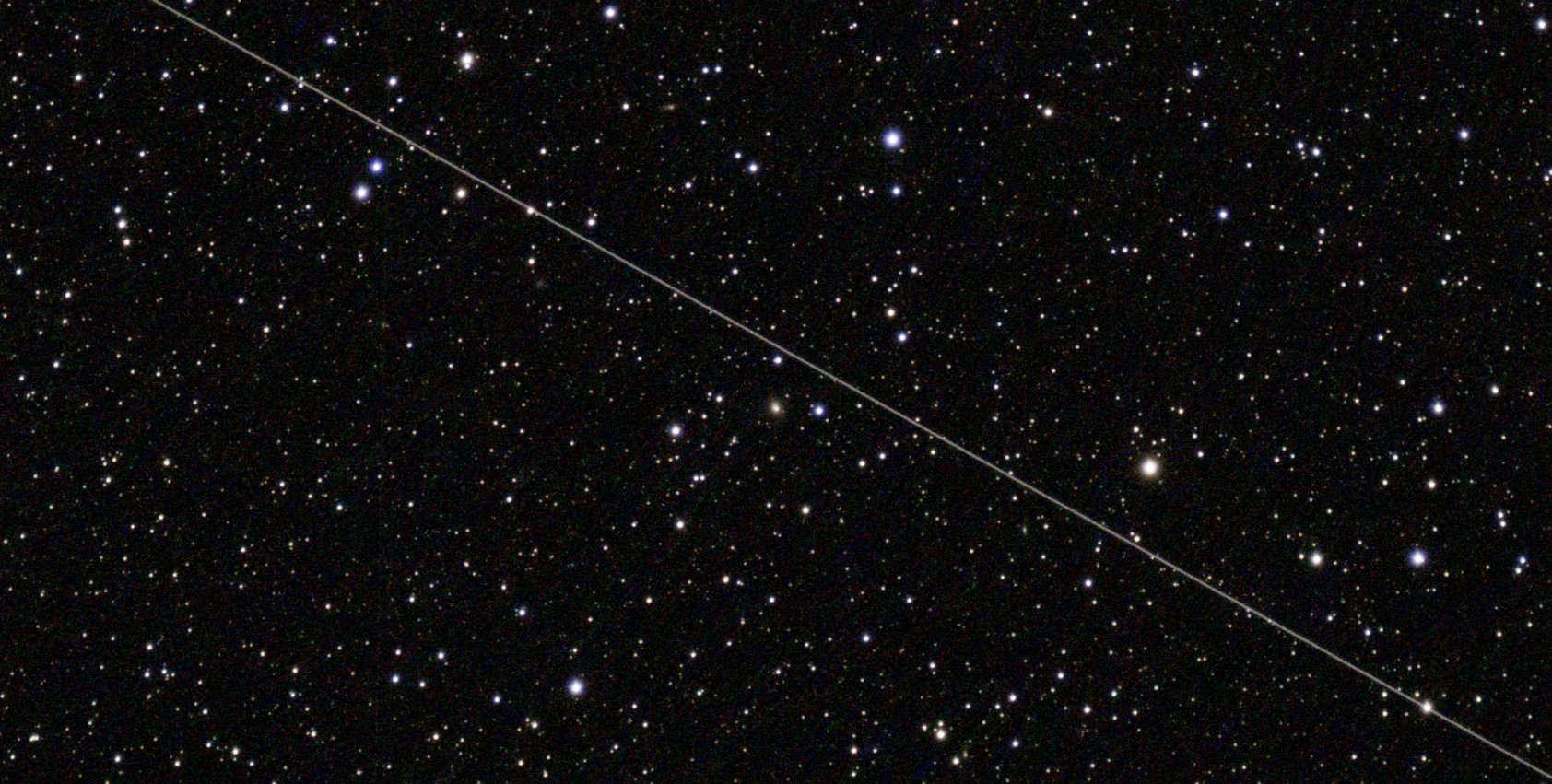}
	\caption{Observation of NGC6482 by the authors with a smart telescope: a streak has completely damaged the image during acquisition .} 
	\label{fig:streak}
\end{figure}

Observational astronomy and astrophotography are challenging, whether professionally or as a hobby, and one of them is the following  the proliferation of mega-constellation satellites such as SpaceX's Starlink, OneWeb and Amazon's Project Kuiper in Low-Earth Orbit (LEO) \cite{hainaut2024contamination,langston2024evaluating}. 
While these satellite networks have started to revolutionize the global internet communications, they also raise growing concerns, for multiple aspects (environment, defence, culture, etc.) \cite{venkatesan2020impact}. 
In particular, the impact of these mega-constellations on astronomy and astrophotography has become a key issue \cite{walker2020impact,varela2023increasing}, calling into question the possibility of observing the night sky without disturbance.
This problem is satellite light pollution, which occurs when orbiting satellites reflect the sun's light unto the Earth \cite{hearnshaw2024light}.
It can make astronomical observations and astrophotography considerably more difficult \cite{lawler2023bright}, and diminishes the quality of images captured by both amateur and professional instruments:
\begin{itemize}
\item Loss of quality: mega-constellation satellites can create light trails as they pass in view of a telescope or camera lens during long exposure photography (\figurename~\ref{fig:streak}). These streaks can compromise image quality by leaving unwanted lines of light across astronomical images from an aesthetic point of view, but also in terms of data loss.
\item Increased sky brightness: the sun's reflection of satellite surfaces can contribute to a general increase in the brightness of the night sky. This makes it more difficult to observe and capture faint, distant celestial objects, such as galaxies, nebulae and faint stars.
\item Reduced contrast: the presence of moving satellites can reduce the contrast between celestial objects and the sky background. Subtle details in astronomical pictures, which depend on a dark, uniformly illuminated night sky, can be compromised by bright streaks and scattered spots.
\item Complications for image calibration: satellite light steaks can disrupt the process by introducing non-stellar elements into the images, making treatment difficult or even impossible.
\end{itemize}

These problems require advanced post-processing to solve, such as technical adjustments and specialized
algorithms like detection, denoising and \textit{inpainting} to mitigate the undesirable effects caused by satellites.

Streaks are also for professional ground-based observatories \cite{hainaut2020impact}, making it imperative to set up a process to estimate the concrete impact on the quality of large digital sky surveys \cite{lu2024impact}, avoid disturbances and correct data if possible \cite{tyson2020mitigation}.
The effect is significant: as recent studies have shown \cite{lawler2023bright,barentine2023aggregate}, the increased traffic in LEO will inevitably lead to a loss of astronomical data and therefore reduce the possibility of discoveries, as weak astrophysical signals are increasingly lost in the noise.
The International Astronomical Union (IAU) recently published a \textit{'Call to Protect the Dark and Quiet Sky from Harmful Interference by Satellite Constellations'} \footnote{\url{https://cps.iau.org/documents/49/techdoc102.pdf}}.
Furthermore, it's a hot topic for space-based observatories like Hubble \cite{kruk2023impact}, which adds a number of constraints that are difficult to resolve, especially given the cost of operating such facilities in space.

Satellites have had and will continue to have a significant influence on this particular phenomenon of light pollution. 
However, it is also important to mention that other fast-moving Near-Earth Objects (NEO), such as meteors and large space debris, or even cosmic rays, have the potential to leave linear features on astronomical images \cite{nir2018optimal}.

In this article, we propose the study of a dataset containing raw astronomical images captured over a year with a smart telescope, in conditions accessible to amateurs, and we evaluate the quantity of images effectively impacted by streaks.
The rest of the paper is structured as follows. 
In Section \ref{sec:related}, we present a brief review of the state of the art concerning the detection of streaks in astronomical images. 
Section \ref{sec:data} describes the dataset of images with smart telescopes between March 2022 and February 2023, while Section \ref{sec:method} proposes a study of this dataset using different methods. 
We discuss the results in Section \ref{sec:discussions}, before concluding and proposing some perspectives in Section \ref{sec:conclusion}.

\section{Related works}
\label{sec:related}

For many years now, the scientific community has proposed many techniques to detect and track satellites, and to deal with the trails they cause in astronomical images \cite{nir2018optimal,calvi2021machine,jiang2023finding}.
Specialized software for astronomical images processing (like SharpCap \footnote{\url{https://www.sharpcap.co.uk}})  provide basic features to manage this problem.

In the Python software ecosystem, we can mention these tools:
\begin{itemize}
\item ASTRiDE aims to detect streaks in astronomical images \cite{2016ascl.soft05009K} with boundary-tracing and morphological parameters. ASTRiDE can detect not only long streaks, but also relatively faint, short or curved ones. As we will see later in this article, this is also a problem because it tends to confuse real streaks with tracking problems -- so it requires a fine configuration like in \cite{duarte2023space}.
\item Authors of \cite{danarianto2022prototype} proposed a Python pipeline for lightweight streak detection, identification and initial orbit determination from FITS raw files captured by amateur-grade telescopes -- but it was tested on only a few images captured around the celestial equator (85). FITS (Flexible Image Transport System) is a file format most commonly used into store astronomical data. 
\item A research team applied the probabilistic Hough transform through a Python scripts using well-known open source libraries like openCV and scikit-images, and by using GPU-specific computation to detect streaks in FITS files captured by the Tomo-e Gozen camera at Kiso Observatory in Japan \cite{cegarra2021streaks}. Unfortunately, the source code is not available.
\item pyradon is a Python package based on Fast Radon Transform (FRT) to find streaks in 2D astronomical images \cite{nir2018optimal}.
\end{itemize}

Some existing approaches are based on Deep Learning.
As an example, an approach based on YOLO (\textit{You Only Look Once}, version 2) was proposed by \cite{varela2019streak} to detect streaks in images captured by a multi-camera system with a wide field of view.
An other YOLO model (version 5) was proposed by \cite{calvi2021machine} to detect streaks in synthetic raw images.
Furthermore, a recent work compared two techniques based on Deep Convolutional Neural Networks: an extended feature pyramid network (EFPN) with faster region-based CNNs (Faster R-CNN) and a feature pyramid network (FPN) with Faster R-CNN \cite{elhakiem2023streak}. 
These approaches are elaborated but they were only tested on synthetic data.
The authors of \cite{mastrofini2022yolo} have used YOLO (version 4) and U-Net to track Resident Space Objects from data captured during night observations in Italy (Campo Imperatore, Gran Sasso) .

We can also add that recent work has also combined a Multi-Layer Perceptron and SHapley Additive exPlanations (SHAP) for Space debris streak classification \cite{ntagiou2023space}.

There is not much work based on images taken in conditions and with equipment accessible to both professionals and amateurs, as far as we know.
In this paper, we show how using different Deep Learning approaches can help to look for real streaks in astronomical raw data that we have captured ourselves using smart telescopes.

\section{Data acquisition}
\label{sec:data}

Electronically Assisted Astronomy (EAA) is increasingly applied by amateur and professional astronomers to observe Deep Sky Objects (DSO), i.e. astronomical objects that are not individual stars or Solar System objects, like nebulae, galaxies or clusters. 
By using a dedicated setup (a reflector/refractor, a modern CMOS camera, a motorized mount) and then applying on-the-fly post-processing, EAA enables enhanced images to be generated and displayed on devices such as laptops or smartphones, even in locations with significant light pollution and poor weather conditions. 
The recent years have brought the emergence of smart telescopes, making sky observation more accessible for the public \cite{parisot2022improving}.
Even the scientific community is taking advantage of these instruments to study astronomical events (i.e. asteroids occultations, exoplanets transits, eclipses) \cite{peluso2023unistellar}.

In this context, MILAN Sky Survey is a set of raw images with DSO visible from the Northern Hemisphere (galaxies, stars clusters, nebulae, etc.), collected during 205 observation sessions \cite{parisot2023milan}. 
These images were captured between March 2022 and February 2023 from Luxembourg Greater Region by using a Stellina smart telescope, based on an Extra Low Dispersion doublet with an aperture of 80 mm and a focal length of 400 mm (focal ratio of f/5), a Sony IMX178 CMOS sensor with a resolution of 6.4 million pixels, an alt-azimuth mount and a CLS filter (City Light Suppression).

As the dataset and the data acquisition process are thoroughly described in \cite{parisot2023milan}, here is a short summary:
\begin{itemize}
\item The Stellina was set to its default configuration, with a 10-second exposure time and 20 dB gain per image. These value provide an effective balance for producing good-quality images: increasing the exposure time could introduce \textit{motion blur} \cite{loke2017astronomical}, while higher gain might raise noise level. 
\item For each observation session, the instrument was installed in a dark environment (no direct light) and properly balanced using a bubble level on a stable floor (it's mandatory to ensure a good tracking). 
\item Observation sessions were conducted only when the sky was clear and of reasonable quality. The authors were always present during data capture to deal with any weather-related issues such as wind, cloud, fog, rain, or disturbance from animals.
\end{itemize}

In total, 205 observation sessions and then 50068 16-bits FITS images of resolutions $3096 \times 2080$ were obtained (corresponding to a field of view of approximately 1° × 0.7°). 
As each raw image was obtained with an exposure time of 10 seconds, it represents a total cumulative time of 139 hours, 4 minutes and 40 seconds.

The result is a set of raw, unfiltered, unprocessed images, captured over twelve months of observations when weather conditions allowed.
The question addressed in this paper is how many of these images are damaged by streaks caused by satellites or other NEO.

\section{Method}
\label{sec:method}

Following the work started in \cite{data24}, and by following a methodology introduced in \cite{astronomy3020009}, we have tested and adapted various Deep Learning approaches to detect streaks in raw astronomical images captured with smart telescopes.
As a use-case, we have applied these approaches on the MILAN Sky Survey dataset, to evaluate the count of FITS files containing streaks.

Data storage and pre-processing, training and execution of Deep Learning models was carried out on a platform managed and maintained by the Luxembourg Institute of Science and Technology (LIST) \footnote{\url{https://www.list.lu/en/informatics/}}.
More precisely, the computations were realized on cores with 128 GB RAM (Intel(R) Xeon(R) Silver 4210 CPU @ 2.20GHz) and NVIDIA Tesla V100-PCIE-32GB. 

In practical terms, FITS files here consist of a single header data unit (HDU) comprising two elements: an ASCII header text, but above all the data, i.e. the raw image in the form of a single-channel matrix of 16-bit integers.
For this work, FITS files have been managed in a Python script with the support of the astropy package \cite{astropy2022astropy}.

\subsection{Detection with ASTRiDE}
\label{sec:astride}

Before applying Deep Learning techniques, we have tested ASTRiDE on data presented in Section \ref{sec:data}, by using different settings and filtering approaches (Table \ref{table:res_astride}).
ASTRiDE first pre-processes the image by removing the background using its level and standard deviation before searching for streaks \cite{2016ascl.soft05009K}. 
It then computes the contour map to identify all object borders within the image. 
ASTRiDE then analyzes the morphologies of each object, as determined by the morphological parameters, to differentiate between streaks and stars.

\begin{figure}[h]
	\centering
	\includegraphics[width=.6\linewidth]{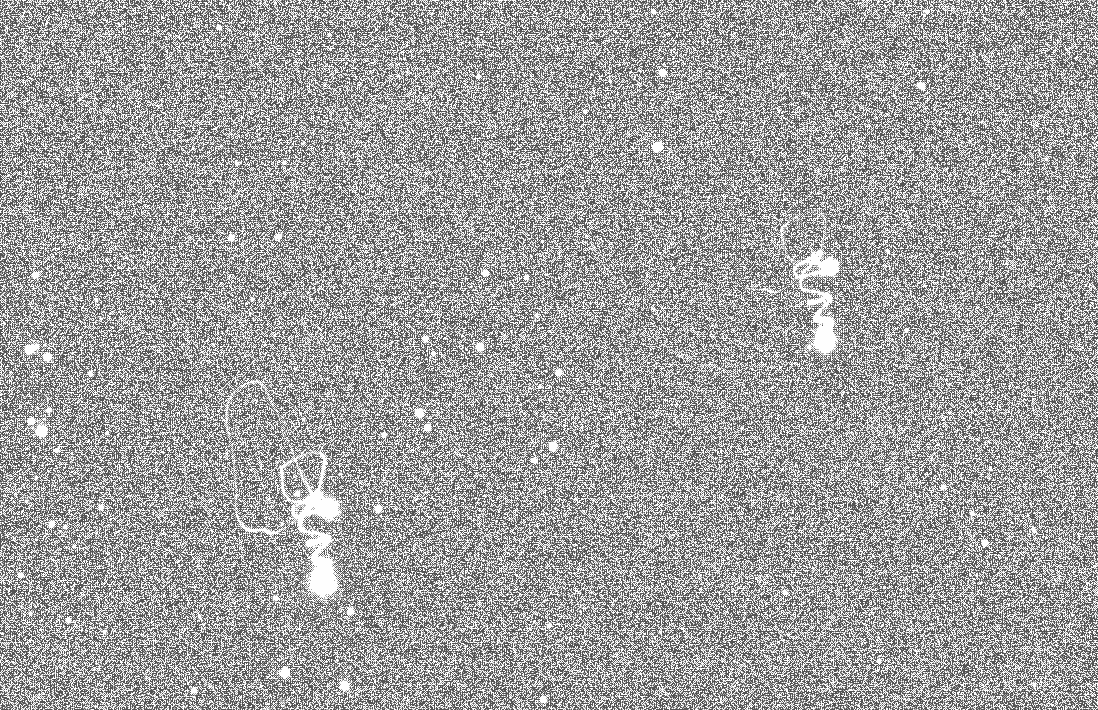}
	\caption{Tracking problem in the FITS file is stored in \textit{NGC457-20220807.zip}: it's not a streak (image from \cite{data24}).} 
	\label{fig:res_astride_falsepositive}
\end{figure}

In practice, the use of ASTRiDE requires the implementation of a Python script based on the dedicated package \footnote{\url{https://pypi.org/project/astride/}}:

\begin{verbatim}
from astride import Streak
streak=Streak(INPUT_PATH,
              contour_threshold=?,
              shape_cut=?,
              output_path=OUTPUT_PATH)
streak.detect()
results=streak.streaks
\end{verbatim}

With this code and by using the default parameters (\textit{contour threshold}=3, \textit{shape cut}=0.2), and by considering FITS images with at least one detected streak, we observed this selection is too large, retaining images with just blurred stars due to tracking problems (example: Figure \ref{fig:res_astride_falsepositive}).

To make a more restrictive selection, we have filtered FITS images with at least one streak with a minimal perimeter (128 pixels), and we have found that streaks are detected in 1316 FITS files for a total of 50068 files, i.e. 2.6 \%. 
In other words, it detected that 137 observation sessions are impacted for a total of 205, i.e. 66 \%.

\begin{figure}[h]
	\centering
	\includegraphics[width=.9\linewidth]{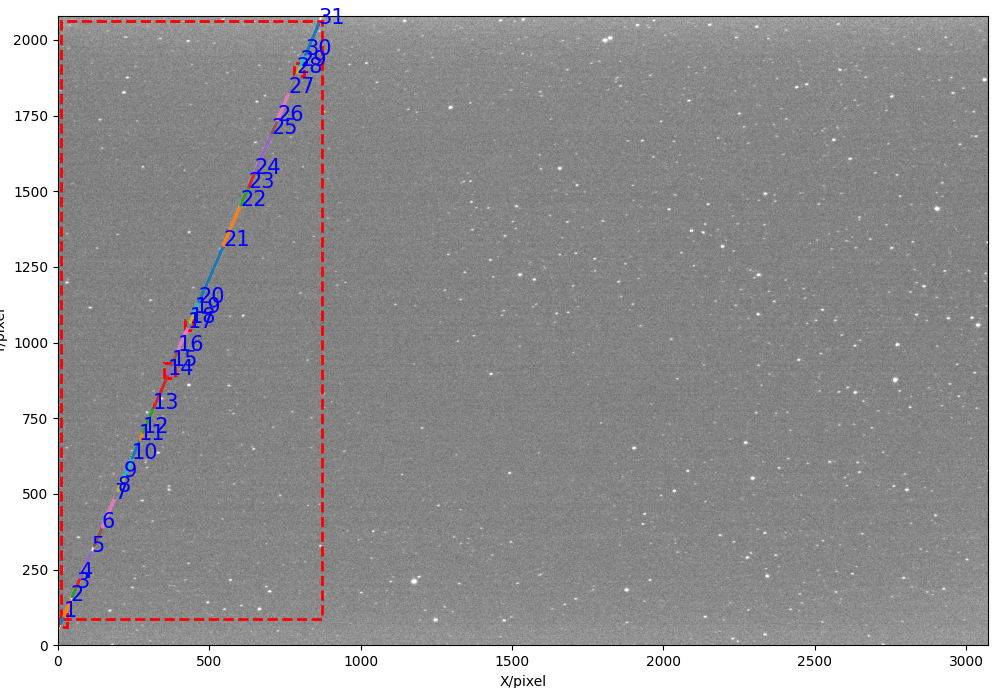}
	\caption{Graphical output of ASTRiDE, after the detection of a large streak in a FITS file captured during an observation session of the Barnard 142 Dark nebula (stored in \textit{Barnard142-143-20220922.zip}, image from \cite{data24}).} 
	\label{fig:res_astride}
\end{figure}

\begin{table*}[]
\centering
\caption{Experiments with ASTRiDE to detect streaks in all the FITS raw images contained in the MILAN Sky Survey. Different settings for ASTRiDE have been tested and compared to check the sensitivity of the method.}	
\label{table:res_astride}
\begin{tabular}{|p{1.25in}|p{1in}|p{1in}|p{1.25in}|}
\hline
  ASTRiDE Settings &
  Filter &
  Raw images with detected streaks &
  Observation sessions impacted \\
\hline
\textit{contour threshold}=3, \textit{shape cut}=0.2 & At least one streak  &  8704/50068 & 198/205 \\
\hline
\textit{contour threshold}=3, \textit{shape cut}=0.2 & At least one streak with perimeter higher than 128 pixels & 1316/50068 & 137/205 \\
\hline
\textit{contour threshold}=5, \textit{shape cut}=0.1 & At least one streak & 903/50068 & 101/205 \\
\hline
\textit{contour threshold}=5, \textit{shape cut}=0.1 & At least one streak with perimeter higher than 128 pixels & 404/50068 & 82/205 \\
\hline 
\end{tabular}
\end{table*}

Moreover, and given the fact that FITS files have an high resolution, the computation may be slow (the tool does not use GPUs to speed-up the analysis -- the ASTRiDE' authors advise to deal with parameters to find a good trade-off between accuracy and speed).  

So we have tried with optimized settings which are proposed by ASTRiDE authors on the official source code repository \footnote{\url{https://github.com/dwkim78/ASTRiDE}}: by increasing \textit{contour threshold} and by reducing \textit{shape cut}, we avoided selecting FITS files that are impacted by minor tracking errors.
Nevertheless, the selection was still too large.

One of the advantages of ASTRiDE is the ability of detecting faint streaks, allowing to detect images damaged by bad tracking (wind, unexpected movement of telescopes, etc.).
For this use-case, and taking into account that our FITS raw images are noisy and far from perfect (especially due to tracking problems), this tool is too sensitive and it is difficult to find the configuration that leads to the detection of streaks by ignoring other issues due to the setup (Figure \ref{fig:res_astride_falsepositive})

\subsection{Detection with a dedicated ResNet50 classifier and XRAI}
\label{sec:xai}

We trained a dedicated Deep Learning classifier to detect images with real streaks -- and ignoring defects due to tracking.
As we have seen in the previous sections, there are many tools available for this task, the aim is not to re-invent the wheel. 
Our aim here is rather to have a model that is fully compatible with our input data and its specific characteristics (in particular the fact that it is raw, unfiltered and not debayered).

To this end, we used ASTRiDE to filter FITS images without any streak and/or defect, as ASTRiDE is a sensible and efficient tool for this task.
Starting from these real raw images, we generated a dataset with synthetic streaks to train and evaluate a binary classifier.
In practice, here are the steps followed:
\begin{itemize}
\item From the raw data described in Section \ref{sec:data}, we built a set of 25070 RGB images with 224x224 pixels -- cutting FITS images into patches to get a resolution that fits to ResNet50 models.
\item For each image, we applied a basic stretch to adjust the brightness and contrast to bring out details and make faint structures more visible.
\item Random synthetic streaks have been added on half of the images, then we formed two distinct groups, so as to associate a class with each image: images with and images without streaks (we made sure that each group was balanced -- to have a classifier with good recall). These streaks were generated by drawing random lines, with different thickness, sizes and color intensity. 
\item We made 3 sets: training, validation and test (80\%, 10\%, 10\%).
\item Inpsired by the official documentation of Keras \footnote{\url{https://keras.io/guides/transfer_learning/}}, a dedicated Python prototype was developed to train a ResNet50 model to learn this binary classification. The basic image processing tasks were performed following best practices for optimizing CPU/GPU usage \cite{castro2023landscape}.
\item We have used the following hyper-parameters for training: ADAM optimizer, learning rate of 0.001, 50 epochs, 32 images per batch. We thus obtained a ResNet50 model with an accuracy of 97\% on the validation dataset. Note that other architectures were also tested (such as VGG16 and MobileNetV2), but the results here are largely similar.
\item At the end, we obtained a model with a precision of 0.940, a recall of 0.805 and then a F1-score of 0.867 (Table \ref{table:res_confusionmatrix}).
\end{itemize}

\begin{table*}[]
\centering
\caption{Confusion matrix with the Resnet50 trained model on test set, i.e. real raw images with randomly added synthetic streaks.}	
\label{table:res_confusionmatrix}
\begin{tabular}{|p{2in}|p{1in}|p{1in}|}
\hline 
                                       		   & Detected: NO & Detected: YES \\
\hline 
Raw images without synthetic streak(s) 		   & 1886         & 103            \\
\hline 
Raw images with synthetic streak(s)    		   & 393          & 1619           \\
\hline 
\end{tabular}
\end{table*}

As done in recent works for the industrial \cite{roth2022towards} and health domains \cite{chaddad2023survey}, and to check the robustness of the trained Resnet50 model, we analysed the output with XRAI (\textit{Region-based Image Attribution}) \cite{kapishnikov2019xrai}. 
Frequently used in eXplainable Artificial Intelligence for Computer Vision tasks, XRAI is an incremental method that progressively builds the attribution scores of regions (i.e. the regions of the image that are most important for classification). 
XRAI is built upon Integrated Gradients (IG) which uses a baseline (i.e. an image) to create the attribution map \cite{sundararajan2017axiomatic}. 
The baseline choice is application-dependent, and in our case we operate under the assumption that a black one is appropriated because it corresponds to the sky background, and the attribution maps is calculated according to the XRAI integration path and reduces the attribution scores given to black pixels. 
In practice, we used the Python package \emph{saliency} \footnote{\url{https://pypi.org/project/saliency/}} and analysed the output heatmap. 
To generate a heatmap indicating the attribution regions with the greatest predictive power, we keep only a percentage of the highest XRAI attribution scores here (for instance, 10 \%).

\begin{figure}[h]
	\centering
	\includegraphics[width=.9\linewidth]{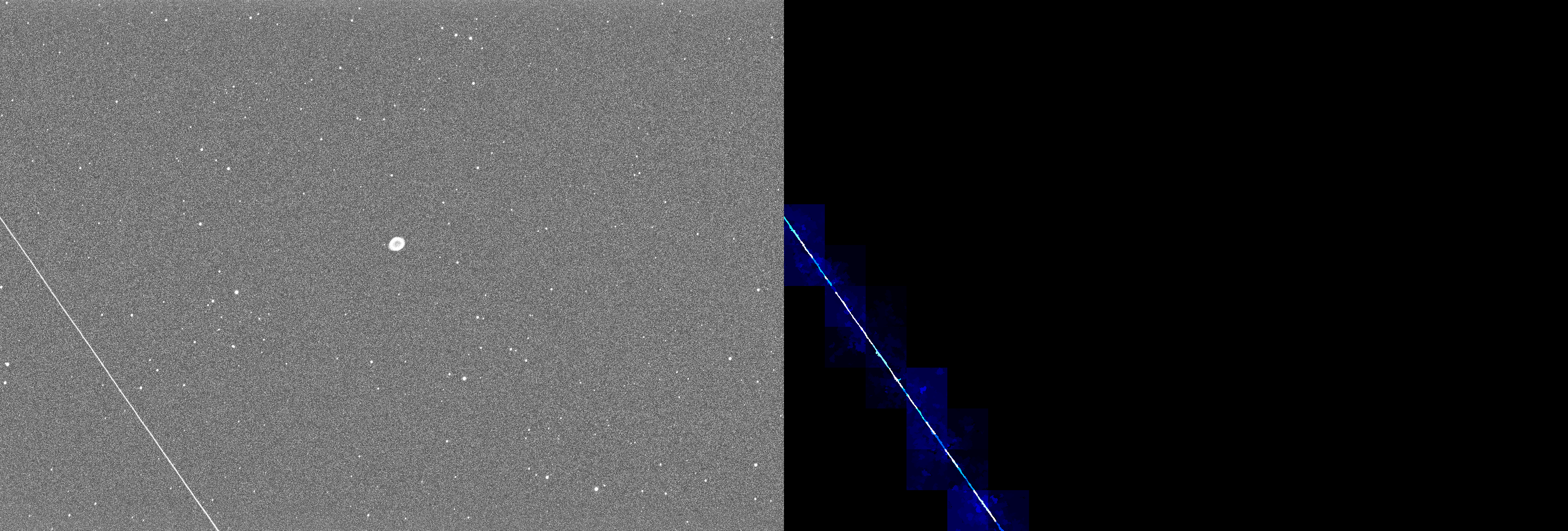}
	\caption{At the top, a stretched raw image of Messier 57 (Ring Nebula) with an exposure time of 10 seconds. At the bottom, the XRAI heatmap highlighting the pixels that are considered by the trained Resnet50 model for the classification, by keeping only 10\% of the highest XRAI attribution scores (image from \cite{data24}).} 
	\label{fig:res_xrai}
\end{figure}

With this pipeline, we have found that streaks are detected in 25 FITS files for a total of 50068 files, i.e. less than 0.05 \%; it detected that 18 observation sessions are impacted for a total of 205, i.e. 0.1 \%.
In this case, we visually noted with the heatmap that the streaks are not caused by tracking problems, but by objects passing through the instrument's field of view during observation.
For example, we can mention the following files present in \cite{parisot2023milan}: image 44 in \textit{Barnard142-143-20220922.zip}, image 22 in \textit{M17-20220723.zip}, image 40 in \textit{M57-20220602.zip}, image 87 in \textit{M65-20220426.zip}, image 40 in \textit{M103-20220808.zip}, image 167 in \textit{M10-20220615.zip}.

\subsection{Fast approximation with a Pix2Pix model}
\label{sec:pix2pix}

Computing a heatmap with XRAI comes at a cost: it requires more computational time and resources than a simple inference of the ResNet50 model.
If we consider the analysis of a $3584 \times 3584$ astronomical image: with no overlap, it may be necessary to evaluate the ResNet50 prediction and the XRAI \textit{heatmap} for 256 patches of $224 \times 224$ pixels -- this may take some time depending on the hardware.
To minimise the number of calculations required, we can apply two simple strategies:
\begin{itemize}
\item Reduce the resolution of the image to decrease the number of patches to be evaluated. In this case, the main problem is the risk of losing details, and in particular of being unable to detect traces that are too fine or too small.
\item Process only a relevant subset of patches -- for example, ignoring those for which the ResNet50 classifier detects nothing.
\end{itemize}

An other solution consists in estimating the XRAI heatmap with Generative Adversarial Networks (GAN), a class of Deep Learning frameworks that are frequently applied for Computer Vision tasks. 
A GAN is composed of two Deep Learning models: a generator that ingests an image as input and provides another image as output, and a discriminator which guides the generator during the training by distinguishing real and generated images. 
Both are trained together through a supervised process -- with the goal to obtain a generator that produces realistic images.
Among the numerous existing GAN architectures, we selected Pix2Pix -- a conditional adversarial approach that was designed for image to image translation \cite{isola2017image}, and applied in many use-cases such as image colouration and enhancement \cite{kumarsingh2023enhanced}.
 
Thus, a Pix2Pix model has been designed to learn the transformation from images with synthetic streaks (like in Section \ref{sec:xai}) and images with the same synthetic streaks but with an other color.
We applied the standard Pix2Pix architecture as described and implemented in the official Tensorflow documentation \footnote{\url{https://www.tensorflow.org/tutorials/generative/pix2pix}}, taking input images of 256 $\times$ 256 pixels, with the same resolution for outputs.
The loss function was based on the Peak Signal-to-Noise Ratio (PSNR), and we trained the model during 100 epochs, the batch size was set to 1, and the process was realized with a learning rate of 0.0001.
To improve the training phase, as described in \cite{tran2021data}, we applied random data augmentation during each epoch with the imgaug Python package \footnote{\url{https://imgaug.readthedocs.io/en/latest/}}.

\begin{figure}[!h]
        \center{\includegraphics[width=.5\linewidth]{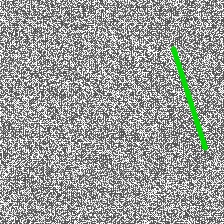}}
        \caption{Example of a 256 $\times$ 256 output generated with a Pix2Pix model from a patch extracted from a raw image -- with highlighted streaks (image from \cite{data24}).}
        \label{fig:pix2pix}
\end{figure}

It led to a Pix2Pix model with a good PSNR (higher that 38.5) -- able to reproduce an annotated image (Figure \ref{fig:pix2pix}), similar to what can be obtained with ResNet50 and the XRAI heatmap.
We simply note that this model is slightly more sensitive to noise, especially if it is grouped in zones (and this can sometimes happen with \textit{hot pixels} \cite{o2023computer}).

In terms of performances, running an inference with the Pix2Pix model on a patch of 256 $\times$ 256 pixels is a better alternative to calculating a heatmap with XRAI on a patch of 224 $\times$ 224 pixels: for example, execution time is halved on a laptop without a GPU.

In practice, we used this Pix2Px model to visually check the results obtained in the previous section, by generating and then viewing the output of each image in which a streak was detected.

\subsection{An annotated dataset and a dedicated YOLOv7 model}
\label{sec:yolo}

As shown in Section \ref{sec:related}, YOLO is a good technique for detecting streaks in images because of its fast, accurate object detection capabilities, making it effective for identifying linear or elongated features even in complex or noisy backgrounds.
One problem described in the literature is that YOLO-based techniques have difficulty detecting objects of small apparent size \cite{wei2024review}. 
In this use-case, this aspect is not important because when the streaks are small (i.e. few pixels), the impact remains limited on the quality of the raw images concerned.

Thus, we have prepared data and model based on YOLOv7 (i.e. YOLO version 7).

\begin{figure}[!h]
        \center{\includegraphics[width=.9\linewidth]{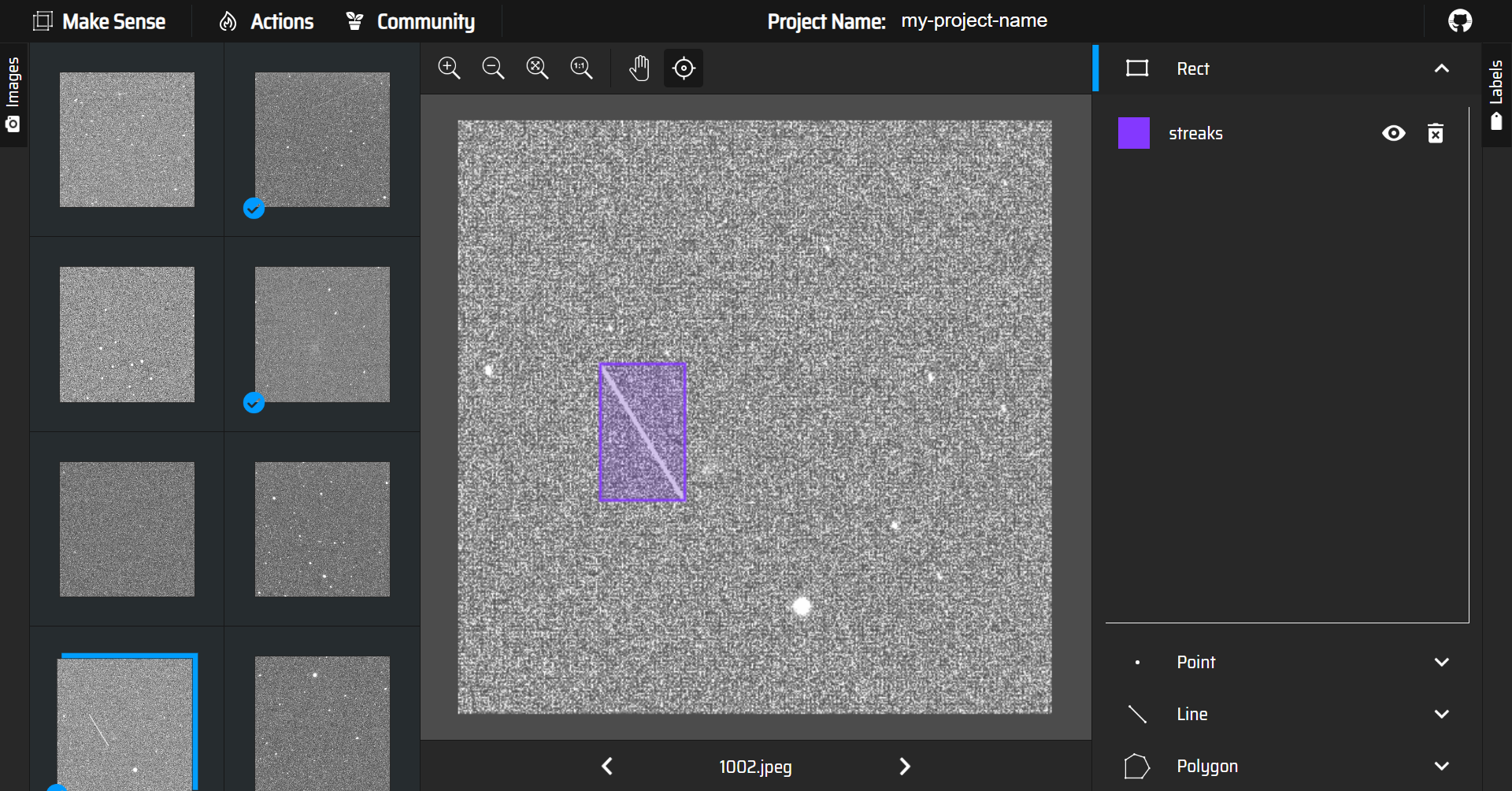}}
        \caption{Adding streaks annotation with the MakeSense software on $640 \times 640$ patches extracted from the MILAN Sky Survey (Section \ref{sec:data}).}
        \label{fig:makesense}
\end{figure}

Similarly to what was shown in Section \ref{sec:xai}, we filtered FITS images with ASTRiDE to keep only images without any streak and/or defect, and we generated a dataset under the YOLO format:
\begin{itemize}
\item From the raw data described in Section \ref{sec:data}, we built a set of 1000 patches with $640 \times 640$ pixels -- by splitting filtered FITS images.
\item For each image, we applied a basic stretch to adjust the brightness and contrast for data augmentation.
\item Random synthetic streaks have been added with a Python script on half of the patches: again, these streaks were generated by drawing random lines, with different thickness, sizes and color intensity. Bounding boxes have been added in separated files with the same name that the image (\textit{IMAGE.txt} for \textit{IMAGE.jpg}, for instance).
\item We then checked these patches by using MakeSense \footnote{\url{https://www.makesense.ai}}: dedicated to annotating data sets for recognition purposes, this interactive web application allows anyone to draw, move, resize and label bounding boxes corresponding to elements present in images (\figurename~\ref{fig:makesense}).
\item The annotated dataset was published under the name StreaksYoloDataset on the Zenodo platform \cite{parisot_2024_14047944}.
\end{itemize}

Based on the official YOLOv7 Python implementation available on Github \cite{wang2022yolov7}, the following command can be used to train during 400 epochs a YOLOv7 model using standard architecture (37M parameters) and transfer learning (based on a pre-trained YOLOv7 model \footnote{\url{https://github.com/WongKinYiu/yolov7/releases/download/v0.1/yolov7.pt}}) with the following parameters:

\begin{verbatim}
python3 train.py --img 640 --weights yolov7.pt
          --data 'data/custom.yaml' --single-cls
          --hyp data/hyp.scratch.p5.yaml
          --cfg cfg/training/yolov7.yaml
          --workers 8 --batch-size 4 --epochs 400
          --name streaks-yolov7
\end{verbatim}

\begin{figure}[!h]
        \center{\includegraphics[width=.5\linewidth]{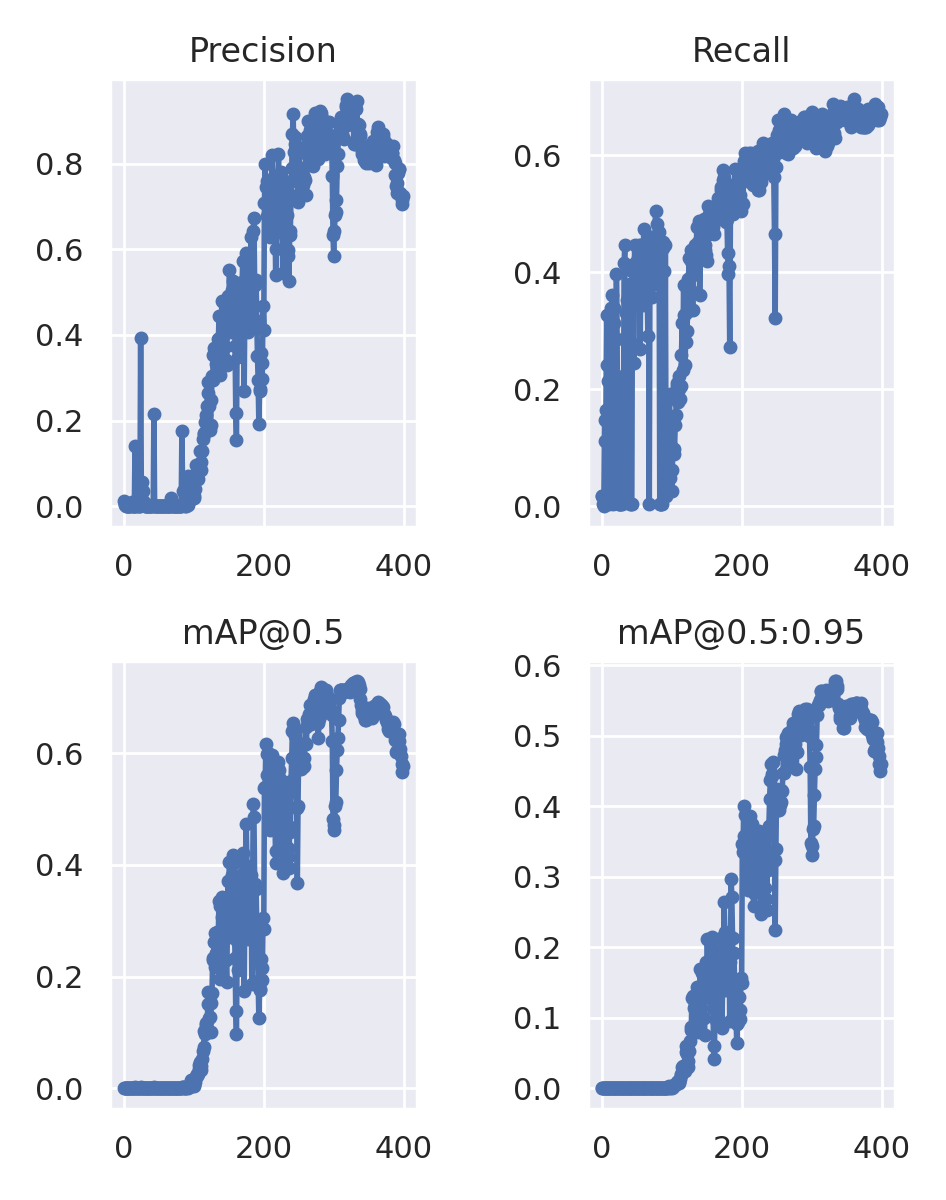}}
        \caption{Evolution of precision, recall and mAP during the training of the YOLOv7 model (37M parameters) during 400 epochs.}
        \label{fig:yolov7_results}
\end{figure}

\begin{figure}[!h]
        \center{\includegraphics[width=.5\linewidth]{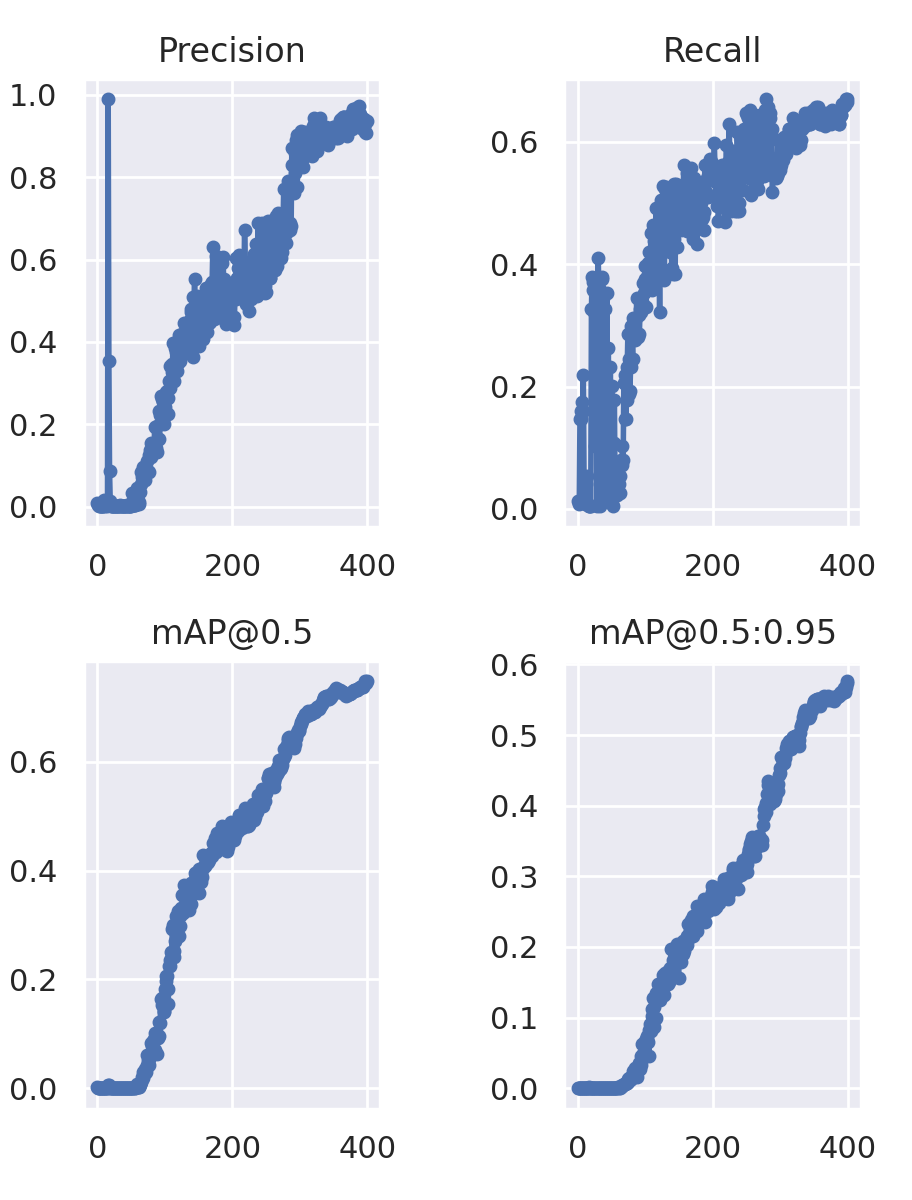}}
        \caption{Evolution of precision, recall and mAP during the training of the YOLOv7 tiny model (6M parameters) during 400 epochs.}
        \label{fig:yolov7_tiny_results}
\end{figure}

As a result, we have obtained a YOLOv7 model with a good accuracy on real images with synthetic streaks (\figurename~\ref{fig:yolov7_results}): on the test set (333 patches), we have a precision of 0.84 and a recall of 0.68.
We have tested an other version of YOLOv7 (tiny architecture with 6M parameters) (\figurename~\ref{fig:yolov7_tiny_results}): on the same test set, we have a precision of 0.96 and a recall of 0.63.

\begin{figure}[!h]
        \center{\includegraphics[width=.9\linewidth]{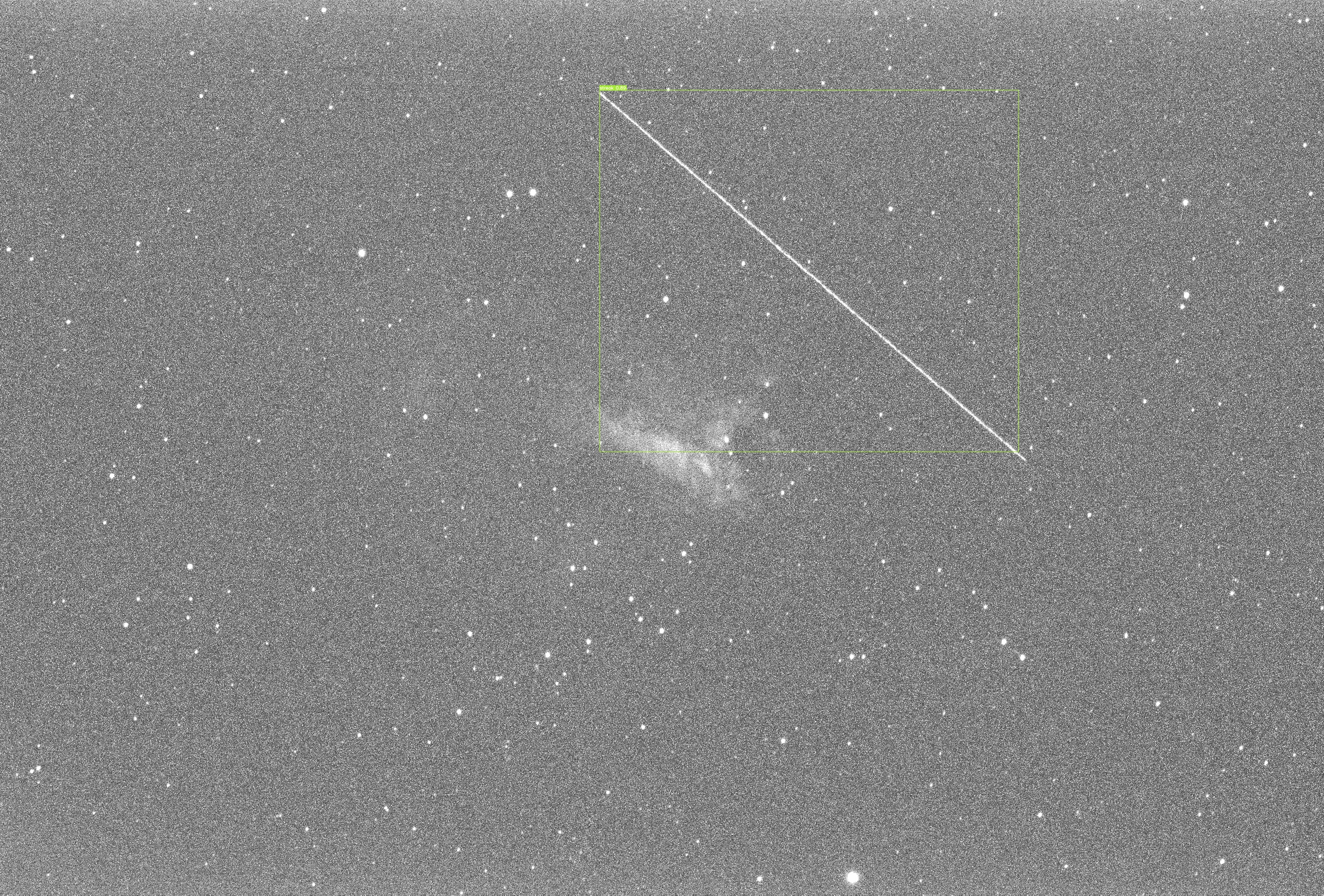}}
        \caption{Output of the YOLOv7 model on a raw FITS image contained in \textit{M17-20220723.zip} (Messier 17 aka Omega Nebula).}
        \label{fig:yolov7_real}
\end{figure}

So we have executed the model on the 50068 FITS files of the MILAN Sky Survey dataset (Section \ref{sec:data}), with the default confidence level of the YOLOv7 reference implementation (0.45):
\begin{itemize}
\item At least a real streak is detected in 89 files with the YOLOv7 model (\figurename~\ref{fig:yolov7_real}).
\item At least a real streak is detected in 85 files with the YOLOv7 tiny model.
\end{itemize}

For the tests, we were able to take advantage of one of the major benefits of YOLOv7, namely the speed of high-resolution image processing (a few tens of milliseconds per image) -- enabling us to process the 50068 images fairly quickly.

\section{Discussions}
\label{sec:discussions}

\subsection{Results}

We have tested different detection methods, and here are our observations:
\begin{itemize}
\item For the FITS files contained in the MILAN Sky Survey, ASTRiDE can detect image without streak with a good precision. As it is difficult to find a good configuration to avoid false positive due to tracking errors, we have a too high selection of images with streak (1316).
\item The Resnet50 model selects 25 FITS files for the \textit{with streak} class, which are all correct after visual inspection, but the selection is not complete, there is here is a too high level of false negative.
\item The YOLOv7 models make a selection of 85 and 89 FITS files containing at least a streak. After an other image-by-image inspection of this list, we can say that 83 files really contain streak.
\item The Pix2Pix model is not a detection model in itself, since it requires the input image to be compared with the output image to see if there is a difference.
\end{itemize}

With these results, and in particular thanks to the YOLOv7 models, we adjust our previous estimate published in \cite{data24} that around 0.16 \% of FITS files contain at least one streak.

\subsection{Limitations}

As it is impossible (and undesirable) to analyse all the raw images by visual inspection, we have used various automated methods to find images potentially affected by streaks. 
These methods may fail to identify some cases (false negatives) if the raw images are of very poor quality, particularly due to tracking problems leading to \textit{motion blur}.

Furthermore, the different approaches have been tested on raw images obtained with specific equipment (80mm aperture, 400mm focal length, recent CMOS sensors, alt-azimuth mount), specific configuration (gain and exposure time) and imperfect weather conditions.
They can therefore be applied to images obtained with identical or similar equipment (i.e. other models of smart telescopes with similar technical characteristics).
Conversely, the application of these techniques to images obtained with instruments of smaller or larger focal length would require the creation of a training set containing this type of data, in order to retrain the models.

\FloatBarrier

\section{Conclusion and perspectives}
\label{sec:conclusion}

This paper presents various approaches based on Deep Learning to detect streaks from astronomical raw images captured with smart telescopes and in conditions accessible for both amateur and professional astronomers.

One approach consists in using ASTRiDE, a reference tool for this task: in this case, it is efficient to detect images without streak.
The second one is a pipeline combining a ResNet50 Deep Learning binary classifier and the XRAI method, allowing the detection of real streaks with a good accuracy.
The third one is an experimental Generative AI model to highlight the pixels corresponding to the detected streaks.
The last one is a YOLOv7 model allowing to detect and estimate the position of streaks, and the associated annotated set of images as been published as an open dataset.

These approaches were tested on raw images captured in the Greater Luxembourg Region over a period of one year and 205 observation sessions.
As a result, we observed that approximatively of 0.16 percent of data are damaged by streaks, potentially caused by satellites, meteors, large debris or cosmic rays.
In light of the projected increase in the number of satellites, it is reasonable to think that this proportion will be revised upwards in the future.

Thus, we plan to collect more data in next months (if possible in both Hemispheres).
Moreover, we plan to work on optimizations to embed the presented Deep Learning approaches into low resource devices, and we will check how they can be applied for Space Domain Awareness. \\

\noindent \textbf{Data availability}: The MILAN Sky Survey (i.e. raw smart telescopes images) can be accessed by following the links listed in \cite{parisot2023milan}. StreaksYoloDataset (i.e. the annotated dataset used to train and evaluate the YOLO models) is available on Zenodo \cite{parisot_2024_14047944}. Additional materials used to support the findings of this study may be available from the corresponding author upon request. \\

\noindent \textbf{Acknowledgements}: This research was funded by the Luxembourg Institute of Science and Technology (LIST), internal grant. Tests were realized on the AIDA platform operated by the Luxembourg Institute of Science and Technology (LIST), thanks to Raynald Jadoul and Jean-François Merche.



\bibliographystyle{splncs04}
\bibliography{refs}

\end{document}